\documentclass[pra,singlespace]{revtex4}
\usepackage{amsmath}
\usepackage{subfigure}
\usepackage{graphicx}

\begin{document}

\title{Creating very slow optical gap solitons with inter-fiber coupling}
 
\author{R. Shnaiderman, Richard S. Tasgal, and Y.B. Band}

\address{Departments of Chemistry and Electro-Optics and
the Ilse Katz Center for Nano-Science, \\
Ben-Gurion University of the Negev,
Beer-Sheva 84105,
Israel}

\begin{abstract}
We show that gap-acoustic solitons, i.e., optical gap solitons with
electrostrictive coupling to sound modes, can be produced with
velocities down to less than 2.5\% of the speed of light using a fiber
Bragg grating that is linearly coupled to a non-Bragg fiber over a
finite domain.  Forward- and backward-moving light pulses in the
non-Bragg fiber that reach the coupling region simultaneously couple
into the Bragg fiber and form a moving soliton, which then propagates
beyond the coupling region.
\end{abstract}


\maketitle

\begin{figure}[ht]
\centering
\includegraphics[width=\textwidth]{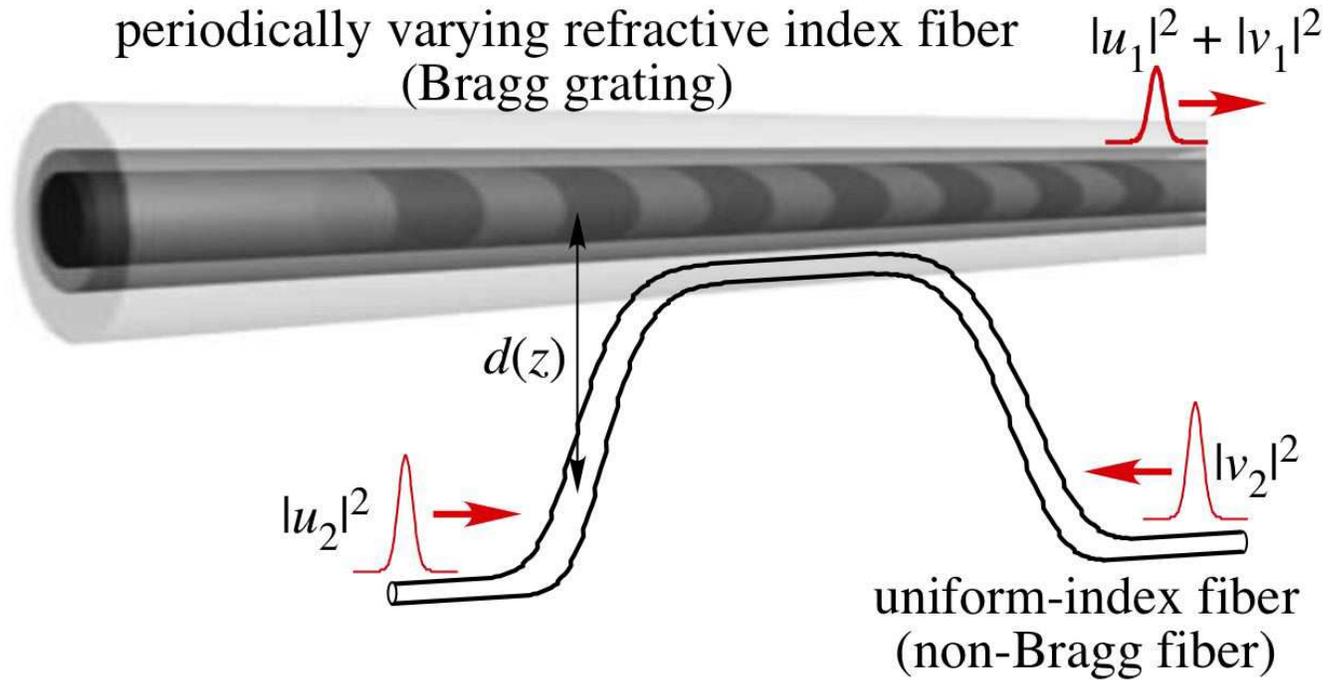}
\caption{(Color online) Schematic drawing of a Bragg fiber and a
non-Bragg fiber that are brought close together so that their
evanescent waves overlap and the light couples between the fibers.
Light pulses are sent into both ends of the non-Bragg fiber, and reach
the coupling region at the same time.  The shape and intensity of the
pulses are adjusted such that they create a gap-acoustic soliton in
the Bragg fiber, along with additional non-soliton radiation.}
\label{Fig.Shnaiderman.schematic}
\end{figure}
\begin{figure}[ht]
\centering
\includegraphics[width=\textwidth]{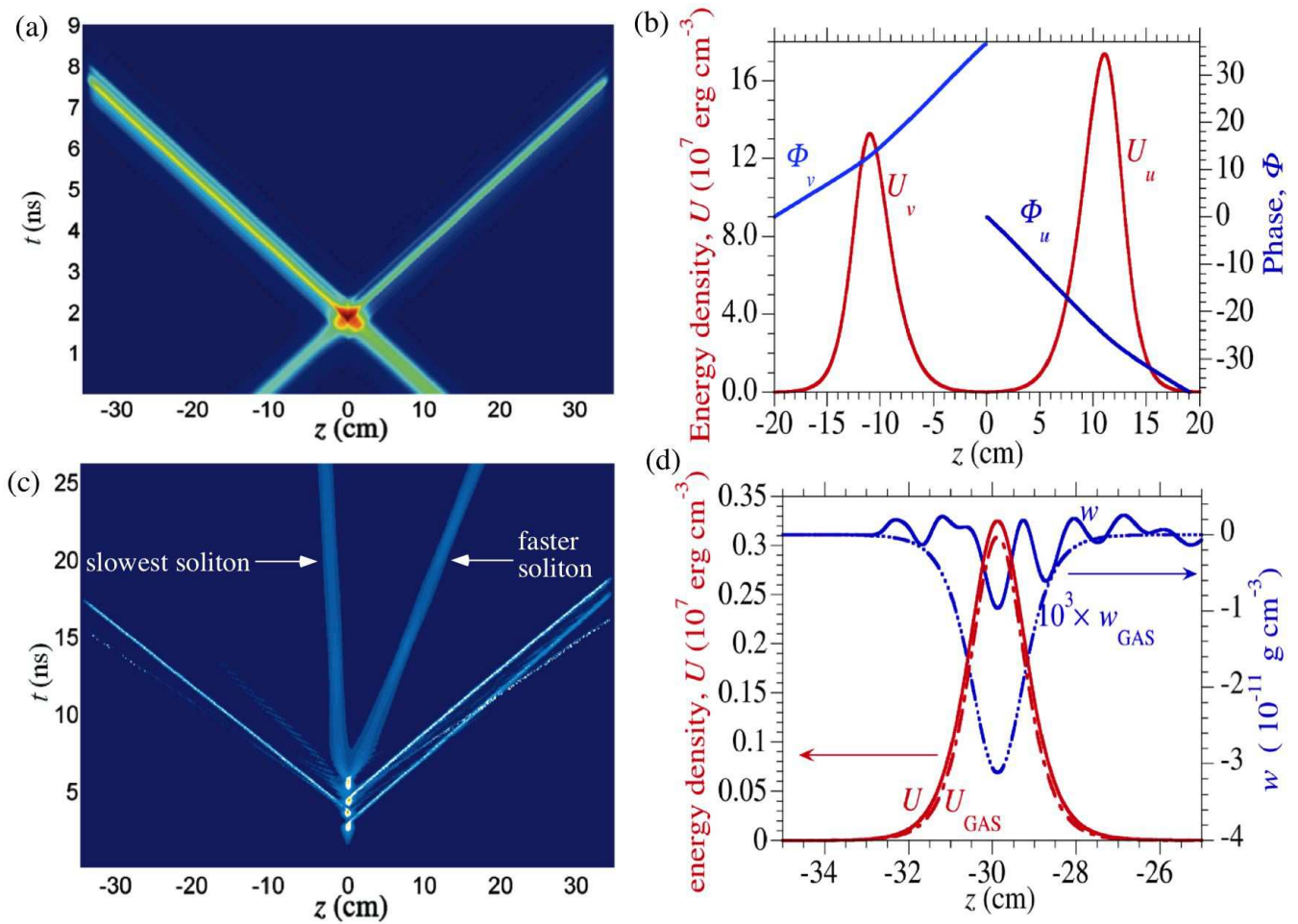}
\caption{(Color online) (a) Surface plot of the energy density in the
non-Bragg fiber, as a function of distance $z$ and time $t$.  (b)
Initial conditions for the simulation showing the energy densities and
phases of the light pulses in the non-Bragg fiber.  (c) Surface plot
of the energy density in the Bragg fiber propagated for $25$\,ns.  The
interactions occur over the first $4$\,ns because of the width of the
input pulses.  The output consists of one very slow GAS (velocity
$0.0246$\,c), and the other pulses are dispersive radiation or faster
GASs.  (d) Light energy density and material density at $185$\,ns
(solid curves), and the exact GAS (dashed curves), with the acoustic
wave of the exact GAS scaled by a factor of $10^3$ to be visible.  The
density in the simulation swamps the soliton wave density.}
\label{Fig.GAS_coupling.simulation}
\end{figure}

There is great interest in slow light \cite{SlowLight.reviews}.
Spectacular slow light results have been achieved in Bose-Einstein
condensates \cite{SlowLight.Hau}, but realizations in room temperature
solid state materials are desirable for many applications.  Optical
fiber Bragg gratings (FBGs) can support one such slow light structure,
the gap-acoustic soliton (GAS), i.e., an optical gap soliton with
electrostrictive coupling to sound waves, which can have velocities
from zero up to the group velocity in the medium \cite{ChenMills.1987,
ChristodoulidesJoseph.1989, AcevesWabnitz.1989, Mok.2006,
TasgalBandMalomed.2007}.  Gap solitons have been produced in the lab
with velocities as slow as 16\% of the speed of light \cite{Mok.2006},
but to date not slower.  Suitable experimental media for GAS
propagation are available, but it can be difficult to create the
correct initial and/or boundary conditions to obtain a GAS in the
first place.  Methods proposed for producing slow GASs include: (1)
fibers with gradually varying modulation depth (apodized)
\cite{MakMalomedChu.apodization.2004}, as was employed in
Ref.~\cite{Mok.2006}, (2) colliding faster moving gap solitons with
each other \cite{MakMalomedChu.collision.2003} or with fiber defects
\cite{MakMalomedChu.JOSAB.2003, ChenMalomedChu.2005}, and (3) growing
a soliton in-place with either distributed \cite{WinfulPerlin.2000} or
localized amplification \cite{MakMalomedChu.localized_gain.2003}.
We propose a method to produce GASs using a FBG and a non-Bragg fiber
that are coupled over a finite distance, as illustrated in
Fig.~\ref{Fig.Shnaiderman.schematic}.  We show that if light pulses,
specially designed by reverse engineering, are sent into the non-Bragg
fiber in the forward- and backward-moving directions such that they
simultaneously reach the inter-fiber coupling region (see
Fig.~\ref{Fig.GAS_coupling.simulation}), a slow GAS can be created in
the FBG. The parameters of the resulting soliton depend on the Bragg
fiber parameters, the coupling to the non-Bragg fiber, and the widths,
intensities, and phases of the input pulses.
This inter-fiber coupling may be contrasted with soliton switching in
uniformly coupled FBGs \cite{HaSukhorukov.GS_switching,
MakMalomedChu.PRE.2004}.  One of the essential differences in this
work is that here the inter-fiber coupling region is finite, and the
light switches exactly once, from the non-Bragg to the Bragg fiber and
then remains in the FBG.

A FBG may be coupled to the non-Bragg fiber by removing some of the
cladding and bringing the fibers close together so that the evanescent
waves of one fiber extend over the core of the other
\cite{Archambault.1994}.  This results in linear coupling between the
light in the two fibers, which is stronger when the fibers are closer
\cite{Orlov.1997,Agrawal.2001}.  The coupling region is finite, and
varies smoothly from the uncoupled to the maximally coupled region.

The dynamics of the system are described by the equations
\begin{subequations}
\label{Eqs:BraggBrillouinKerr}
\begin{eqnarray}
0 & = & i k_0' u_{1,t}
      + i u_{1,z}
      + \kappa v_1
      + \lambda(z) u_2
      + C  
        (|u_1|^2 + 2 |v_1|^2) u_1
      + \chi_{es} w u_1 ,
      \label{Eq:u1} \\
0 & = & i k_0' v_{1,t}
      - i v_{1,z}
      + \kappa u_1
      + \lambda(z) v_2
      + C  
        (2 |u_1|^2 + |v_1|^2) v_1
      + \chi_{es} w v_1 ,
      \label{Eq:v1} \\
0 & = & w_{tt}
      - \beta_s^2 w_{zz}
      - \Gamma w_{tzz}
      + \lambda_{es} (|u_1|^2 + |v_1|^2)_{zz} ,
      \label{Eq:w} \\
0 & = & i k_0' u_{2,t}
      + i u_{2,z}
      + \lambda(z) u_1 ,
        \label{Eq:u2} \\
0 & = & i k_0' v_{2,t}
      - i v_{2,z}
      + \lambda(z) v_1 ,
      \label{Eq:v2}
\end{eqnarray}
\end{subequations}
where $z$ is the spatial coordinate, $t$ is time, $u_1$ is the
amplitude of the envelope of the forward-moving electric field in the
Bragg fiber, $v_1$ is for the corresponding backward-moving field,
$u_2$ and $v_2$ are the fields in the non-Bragg fiber, $w$ is the
material density of the FBG and subscripts in the independent
variables denote differentiation.  The reciprocal of the group
velocity is $k_0' = \frac{d}{d\omega} [n(\omega) \omega /
c]_{\omega_0}$, $\kappa$ is the Bragg scattering coefficient, which
scatters light back into the same fiber, $\lambda(z)$ is the position
dependent inter-fiber coupling coefficient (co-directional;
contra-directional inter-fiber coupling is taken to be zero), $C =
2\pi \omega_0[n(\omega_0) c]^{-1} \, 3 \chi^{(3)} (\omega_0; \omega_0,
-\omega_0, \omega_0)$ is a self-phase modulation coefficient,
$\chi_{es} = (\omega_0/c) \partial n/\partial w$ and $\lambda_{es} =
(2\pi)^{-1} n(\omega_0) w \partial n/\partial w$ are electrostrictive
coefficients, $\beta_s$ is the speed of sound in the fiber, and
$\Gamma$ is the phonon viscosity coefficient.  The inter-fiber
coupling is due to the overlap between the evanescent tails of the
modes in the neighboring fibers \cite{Orlov.1997,Agrawal.2001}.
We have included the coupling of light ($u_1$, $v_1$) to the material
density ($w$) in the Bragg fiber but not for the non-Bragg fiber
($u_2$, $v_2$) because the propagation velocities in the latter (but
not the former) are too fast to be affected.  Kerr effects are
unimportant in the non-Bragg fiber and are neglected.
We used typical physical parameters for bulk fused silica at
wavelength $0.8\,\mu$m \cite{RatForet.2005,Milam.1998}: The index of
refraction is $n_0 = 1.45$, the Kerr coefficient is $C = n_2^I
n(\omega_0) \omega_0 /(2\pi)$, with $n_2^I = 2.8 \times
10^{-16}$\,cm$^2$/W, the material density is $w = 2.2\,$g/cm$^3$, and
dependence of the refractive index on it is $\partial n/\partial w =
0.2$ cm$^3$/g, the speed of sound is $\beta_s = 5.9\,$km/s, the phonon
viscosity is $\Gamma = 6.9 \times 10^{-7}\,$m$^2$/s, and the Bragg
grating coefficient $\kappa$ is proportional to the modulation depth
of the grating; we use $\kappa = 2.7\,$cm$^{-1}$.  The inter-fiber
(codirectional) coupling, due to overlap of the evanescent fields with
the cores of the adjacent fibers, depends on the distance between the
Bragg and non-Bragg fibers.  We take the coupling coefficient to be
$\lambda(z) = A_\mathrm{cpl} \exp [(z/z_\mathrm{cpl})^4]$, with
$A_\mathrm{cpl} = 1\,$cm$^{-1}$, $z_\mathrm{cpl}=0.2\,$cm, i.e., a
super-Gaussian, which is fairly flat in the middle and smoothly but
quickly decreases to zero at the edges.

The inputs of light from the non-Bragg fiber that yield GASs in the
Bragg fiber are determined by reverse engineering: A GAS is propagated
numerically in the Bragg fiber, initially moving toward the region
that is coupled to the non-Bragg fiber.  When we find an instance in
which an appreciable amount of the energy escapes into the non-Bragg
fiber, we run the simulation in the opposite direction, inserting the
obtained light pulses only into the non-Bragg fiber.  The scheme is
optimized further by fine-tuning the pulse parameters.  One
consideration is that the two input pulses must be asymmetrical to
produce a soliton that moves away from the coupling region without
returning the light back to the non-Bragg fiber.  Another is that the
initial light pulses must be stronger than the results of the
time-reversal output pulses, since there is less than complete
coupling of the light between the fibers.  We achieved better results
with stronger but shorter coupling regions.

\begin{figure}[ht]
\centering
\includegraphics[width=\textwidth]{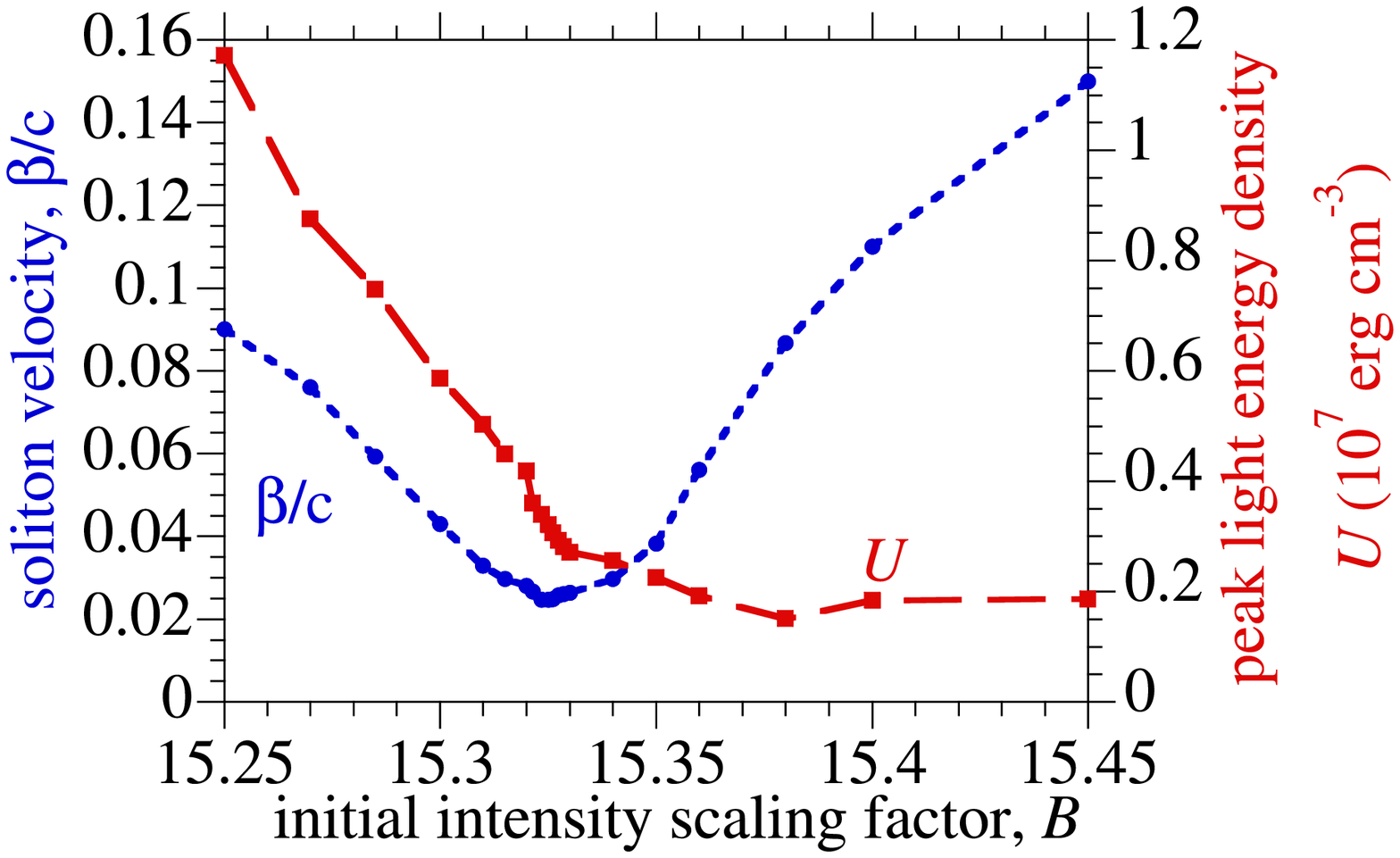}
\caption{(Color online) Plot of the velocity ($\beta$) and peak
energy density ($U$) of the slowest GAS as functions of the initial
intensity scaling factor ($B$), the ratio of the amlpitude of the
input pulses over the amplitude of the output pulses in the
time-reversal simulation.}
\label{Fig.GAS.v_I_vs_B}
\end{figure}

We succeeded in creating GASs in the Bragg fiber with velocities from
close to the group velocity in the medium down to less than $c/40$.
Figure~\ref{Fig.GAS_coupling.simulation} shows an example of such a
simulation.  Light pulses enter the non-Bragg fiber from opposite ends
[panels (a) and (b)], and they reach the region with non-zero
inter-fiber coupling at the same time.  After a finite interaction
time, light propagates into the uncoupled region of the Bragg fiber,
some of it as a very slow GAS (velocity $0.0246$\,$c=0.0357$\,$v_g$),
while other light propagates as faster GASs or as dispersive radiation
[panel (c)].  The energy density and material density of the slowest
GAS is shown in (d).  Figure~\ref{Fig.GAS.v_I_vs_B} shows the
parameters of the GAS produced versus an input intensity factor $B$,
where $B$ is the ratio of the input amplitude in the non-Bragg fiber
divided by the output amplitude in the non-Bragg fiber after the
time-reversed simulation.  This illustrates the optimization we
carried out on the family of results.

Let us review the approximations.
Equations~(\ref{Eqs:BraggBrillouinKerr}) omit self-phase modulation
and low wavenumber acoustic waves in the non-Bragg fiber, and
interactions with high wavenumber acoustic waves (i.e., Brillouin
scattering) in both fibers.
Low wavenumber acoustic waves were shown in \cite{Tasgal.2010} to be
critical for optical gap solitons when the soliton velocity is $\leq
v_g/200$, because the soliton has a momentum minimum near that
velocity.  We did not attain such a slow velocity, but the results
were within an order of magnitude of it, and with better optimization
of the system parameters it is might be possible to achieve a soliton
velocity at which the field $w(z,t)$ is essential.
Brillouin scattering in a waveguide is enhanced by a Bragg fiber,
but---very importantly---only outside the band gap \cite{Ogusu.2000}.
Within the band gap created by a Bragg grating, gap solitons do not
show notable Brillouin back-scattering, and the models work well
without inclusion of Brillouin fields (see  e.g., \cite{Mok.2006}).
Reference \cite{Tasgal.2010} found that under realistic experimental
conditions, there may be significant Brillouin scattering only when
the gap solitons have velocities very close to the group velocity of
light.
In the non-Bragg fiber, low wavenumber acoustic waves (inclusion of
which would amount to a generalized Zakharov system
\cite{Malomed.Zakharov.1997}) will be insignificant for realistic
light pulses because pulses move at the group velocity of light, which
is large.  Brillouin scattering can be important in non-Bragg fibers,
but fiber lengths, especially for short pulses, need to be fairly
long.  The non-Bragg fibers herein need just be long enough to bring
the light to the inter-fiber coupled region, so Brillouin scattering
can be avoided by not using excessively long lead-in fibers.
Within the inter-fiber coupling region, the dynamics are transient,
highly nonlinear, and complex.  The interaction times (a few ns) was
not long enough for acoustic waves to develop much, and the
(transient) high wavenumber acoutic waves that result are not likely
to induce an effective grating consistently in-phase with the light
fields due to the messiness of the fields and non-uniformity of the
fibers [note $\lambda=\lambda(z)$] in the inter-fiber coupled region.
We omitted less important nonlinearities in order to get to he
essence of the problem.  However, that opto-acoustic interactions
might have some effect could not be completely ruled out, so we ran
simulations with all the nonlinearities described here.  We used the
model in \cite{Tasgal.2010}, which is the same as in
\cite{KaiserMaier.1972}, but does not approximate the speed of sound
as zero and does not include equations for temperature dynamics.  We
found only quantitative, not qualitative, modifications of the
dynamics.

The reverse engineering simulations provide a framework for
understanding the physics and they may help to motivate guesses for
experimental parameters.  Ultimately, the reverse engineering
simulations can be dispensed with, and one may take initial pulses as
hyperbolic secants or Gaussians of suitable width, amplitude, and
frequency, and relative position.

In summary, we propose and numerically demonstrate a method for
creating optical gap solitons (or gap-acoustic solitons) in a fiber
Bragg grating by side-coupling the light over a finite region from a
non-Bragg fiber.  Note that this scheme could be used as an
\textit{and} switch, since a soliton will only be produced if there is
input from \textit{both} directions of the non-Bragg fibers.  The
solitons we produced have velocities down to approximately 2.5\% of
the speed of light.  Much below this velocity the solitons are subject
to stronger supersonic instability \cite{TasgalBandMalomed.2007}, and
velocity zero is inaccessible since the solitons need enough velocity
to escape the inter-fiber coupling region.  Combining this with other
techniques, such as colliding solitons
\cite{MakMalomedChu.collision.2003} or making positive use of the
supersonic instability \cite{TasgalBandMalomed.2007} may be used to
create yet slower or even stopped light pulses.

This work was supported in part by grants from the U.S.-Israel
Binational Science Foundation (No.~2006212), the Israel Science
Foundation (No.~29/07), and the James Franck German-Israel Binational
Program.

\end{document}